\documentclass[manuscript]{aastex}

\shorttitle{A Study of the 2014 January 6 Ground Level Enhancement}
\shortauthors{Thakur et al.}

\begin{document}

\title{Ground Level Enhancement in the 2014 January 6 Solar Energetic Particle Event}

\author{N. Thakur\altaffilmark{1,2}, N. Gopalswamy\altaffilmark{1}, H. Xie\altaffilmark{1,2}, P. M\"akel\"a\altaffilmark{1,2}, S. Yashiro\altaffilmark{1,2},\\ S. Akiyama\altaffilmark{1,2}, J.M. Davila\altaffilmark{1}}

\altaffiltext{1}{NASA Goddard Space Flight Center, Greenbelt, Maryland, USA.}
\altaffiltext{2}{The Catholic University of America, Washington DC, USA.}

\begin{abstract}
We present a study of the 2014 January 6 solar energetic particle (SEP) event, which produced a small ground level enhancement (GLE), making it the second GLE of this unusual solar cycle 24. This event was primarily observed by the South Pole neutron monitors  (increase of ${\sim}$2.5\%) whereas a few other neutron monitors recorded smaller increases. The associated coronal mass ejection (CME) originated behind the western limb and had the speed of 1960 km/s. The height of the CME at the start of the associated metric type II radio burst, which indicates the formation of a strong shock, was measured to be 1.61 Rs using a direct image from STEREO-A/EUVI. The CME height at the time of GLE particle release (determined using the South Pole neutron monitor data) was directly measured as 2.96 Rs, from the STEREO-A/COR1 white-light observations. These CME heights are consistent with those obtained for the GLE71, the only other GLE of the current cycle as well as cycle-23 GLEs derived using back-extrapolation. GLE72 is of special interest because it is one of the only two GLEs of cycle 24, one of the two behind-the-limb GLEs and one of the two smallest GLEs of cycles 23 and 24.
\end{abstract}

\keywords{acceleration of particles - shock waves - Sun: coronal mass ejections (CMEs) - Sun: flares, Sun: particle emission - Sun: radio radiation}

\section{Introduction}
The solar energetic particle (SEP) events, where the particles accelerated to GeV energies are subsequently detected on the ground as a result of the air-shower process, are known as the ground level enhancements (GLEs). With a typical of a dozen GLEs per cycle, an average of 16.3\% SEP events were GLEs in cycles 19-23 \citep{Cliver1982, Cliver2006, SheaSmart2008, Mewaldt2012, Nitta2012, Gopalswamy2012a}. In cycle 24 this fraction is much smaller (6.4\%) with 2 GLEs out of 31 large SEP events \citep{Gopalswamy2014}. This also much smaller than the ratio 18\% when the first five years of cycle 23 are considered. GLEs are typically associated with intense flares (median soft X-ray intensity $\sim$X3.8) and fast coronal mass ejections (CMEs; average CME speed $\sim$2000 km/s - see Gopalswamy et al. 2012a).



In cycle 24 only one GLE (2012 May 17 - see Gopalswamy et al. 2013b) occurred prior to the 2014 January 6 GLE (GLE72) studied here. The 2014 January 6 GLE is unique because despite it being a backside event, observations from the Solar Terrestrial Relations Observatory (STEREO) provide a view of the source region and the early CME propagation in full detail. This is the first ever set of observations that covers the spatial domain of interest for a behind-the-limb GLE. We exploit this unprecedented opportunity to understand the early kinematics of the associated CME and shock formation from extreme ultraviolet (EUV) and the height of the CME at the time of particle release.

In GLEs the first arriving particles can be considered relatively scatter-free so they provide information on the solar particle release (SPR) time, i.e., the time when the accelerated particles were injected into interplanetary space. It has been widely accepted that the majority of the GLE particles are accelerated by CME-driven shocks \citep{Reames2013, ReamesTylka2013}. Thus the knowledge of the SPR time together with the coronagraphic observations can be used to determine the height at which the GLE particles are released. For a shock driven by a high-speed CME, the height of the shock can be approximated as the height of the CME itself. Combined measurements from the Large Angle and Spectrometric Coronagraph (LASCO) onboard the Solar and Heliospheric Observatory (SOHO), with a field of view (FOV) of 2.5-32 Rs \citep{Brueckner1995} and the inner coronagraph (COR1) on STEREO with a FOV extending closer to the solar surface (1.4 Rs) \citep{Thompson2003}, provide accurate CME kinematics, the height of shock formation and the height of the shock at the SPR time. Additional details of the eruption, down to the solar surface, are obtained from the Extreme Ultraviolet Imager (EUVI) on STEREO \citep{Wuelser2004, Howard2008}. It is important to understand the GLE of 2014 January 6, not only because it is the only other GLE of the unusually weak cycle 24 but also because there have been nearly 40 large (X-ray magnitude ${\ge}$ M5) SEP events in this cycle out of which eight were well-connected  (source region longitude range: W55-W88) but did not produce a GLE \citep{Gopalswamy2013a}.

\section{Observations and discussion}

As shown in Figure 1(a), the South Pole neutron monitor (SPNM)  \citep{Bieber2013, Evenson2011} recorded the onset of GLE72 around 07:58 UT and measured a maximum rise of ${\sim}$2.5\% above the cosmic-ray background. A few other neutron monitors including McMurdo and Tixie Bay, observed smaller increases (${<}$2\%) making it difficult to determine the anisotropy during GLE72. This event was observed by particle detectors onboard the Geostationary Operational Environmental Satellites (GOES) at higher energies (${>}$ 700 MeV), and by detectors on the Advanced Composition Explorer (ACE), STEREO and Wind spacecraft \citep{Rosenvinge1995} at lower energies (${\le}$ 100 MeV). The GOES 420-510 MeV proton channel observed the onset at ${\sim}$08:12 UT. The increase in intensity of ${>}$700 MeV protons from GOES (Figure 1(b)) confirms the 2014 January 6 event to be a GLE, because all GLEs in cycles 23 and 24 have discernible enhancements in the GOES ${>}$700 MeV energy channel \citep{Gopalswamy2014, TylkaDietrich2009}. Both the GOES-13 and GOES-15  \citep{Rodriguez2014} detected this event with the GOES-13 proton fluxes being greater at higher energies.



The Learmonth radio observatory observed a metric type II  (m-Type II) radio burst at 07:45 UT, which lasted until 08:05 UT  (Figure 1 upper right). The radio and plasma wave receiver, WAVES \citep{Bougeret1995}, on the Wind spacecraft detected the type II burst in decameter-hectometric wavelengths (DH-Type II) from 14 MHz down to 1 MHz and in the kilometric domain (km-Type II) from 1 MHz to 200 kHz (Figure 1 lower right). A type II burst extending from metric to kilometric wavelengths is an indication that a CME is able to drive a strong shock from the low solar corona into the interplanetary medium and is typical of events with GLEs \citep{GopalswamyPlanetaryWorkshop2011, Gopalswamy2010}.

The CME of interest was observed by STEREO-A/COR1, STEREO-A/EUVI, and SOHO/LASCO (Figure 2). STEREO-A (located at W151) confirmed that the CME originated behind the limb (Figure 2(a)). STEREO-A/EUVI, STEREO-A/COR1, and SOHO/LASCO-C2 first observed the CME at 07:30 UT, 07:55 UT, and 08:00 UT, respectively. These multi-spacecraft coronagraphic observations provide excellent data to characterize the CME kinematics. The eruption was also observed by the Atmospheric Imaging Assembly (AIA) on the Solar Dynamics Observatory (SDO) \citep{Pesnell2012, Lemen2012} over the west limb. A C2.2 solar flare starting at 07:30 UT was reported by the on-line Solar Geophysical Data, but this flare class does not represent the true magnitude of the soft X-ray flare because most of the flare structure was occulted by the west limb.

\section{ Onset of the event and calculation of SPR time}

We used the 2-minute data from SPNM to estimate the SPR time. Due to this being a small GLE and some fluctuations in the background intensity just before the onset of the event, we estimate the onset of the GLE to lie between 07:54 UT and 08:02 UT. We take the median of this, 07:58 UT, as the onset time for GLE72 at SPNM. The solar wind speed was ${\sim}$350-400 km/s. The geomagnetic conditions were quiet with Kp index of +1 around the event onset (Geomagnetic data service, Kyoto: http://wdc.kugi.kyoto-u.ac.jp). The asymptotic directions for several neutron monitors are shown in Figure 3. The direction of the interplanetary magnetic field (IMF) fluctuated near the event onset. The particle arrival directions based on the IMF are also shown in Figure 3. We assume that after being accelerated and injected into the interplanetary medium the particles traveled the nominal path length of 1.2 AU before reaching Earth. SPNM has a geomagnetic rigidity cutoff of 0.09 GV (${\sim}$0.004 GeV) but the atmospheric effects dominate the NM measurements. The high altitude of SPNM (2820 m) can allow observation of lower energy particles (${\sim}$300 MeV). However, the first arriving GLE particles would be of the highest energies from the accelerated population. In addition, the fact that GOES recorded onset of $>$700 MeV protons several minutes after the event onset at SPNM also confirms that the GLE72 onset at SPNM was due to particles of energies $>$700 MeV. Thus we take 1.0 GeV as the energy of the first arriving particles at SPNM. The 1.0 GeV protons (speed of ${\sim}$2.63x10${^5}$ km/s) would take 11.4 minutes to travel the distance of 1.2 AU and therefore, the SPR time is estimated to be 07:47 UT. In the CME normalized frame of reference (incorporating ${\sim}$8.3 mins for propogation of solar electromagnetic radiation to Earth) the SPR time is ${\sim}$07:55 UT. This timing is consistent with many previous observations that show that the shock forms several minutes before SPR occurs \citep{Kahler2003, Gopalswamy2012a, Gopalswamy2013a}.


\section{ CME kinematics and height at SPR}

We utilized the observations from STEREO and SOHO/LASCO to obtain the early kinematics of the CME using the graduated cylinder model of a flux-rope \citep{Thernisien2011}. The flux-rope in this model has conical legs with circular cross section, a circular front, and it expands self-similarly. The flux-rope fit to CME observations provides the height-time information and width for the CME as well as the source region coordinates. The height-time, speed, and acceleration for the GLE72 CME are depicted in Figure 2. The fit provided W102 as the source longitude of the flux-rope. The CME reached its peak speed ($\sim$1960/s) and experienced peak acceleration ($\sim$2.08 km/s${^2}$) at 08:00 UT and 07:55 UT respectively, thus the CME  acceleration was at its maximum at the time of SPR. The drag force decelerated the CME and the acceleration ceased between 08:00 and 08:05 UT. The time of shock formation is indicated by the onset of m-Type II burst \citep{Reames2013, Gopalswamy2013b, Gopalswamy2012a, Gopalswamy2012b, Reames2009a, Reames2009b}. Type II onset times provide crucial information about particle acceleration because electromagnetic radiation does not depend on magnetic connectivity. The STEREO-A/EUVI observed the associated eruption at 07:45 UT (Figure 2(a)), at the onset of m-Type II, and hence provides a direct measurement of CME height at shock formation as 1.61 Rs. STEREO-A/COR1 observed the CME at 07:55 UT (Figure 2(c)), thus giving the CME height of 2.96 Rs at SPR.

The timeline of events during the GLE based on the measurements and calculations are shown in Table 1. The CME accelerated from 07:30 UT. After its formation at 07:45 UT, the CME-driven shock traveled for ${\sim}$10 minutes (acceleration time) before particles were released into the interplanetary medium.

\section{Comparison of GLE72 with other GLEs}
We now compare the 2014 January 6 GLE with a few GLEs in cycles 23 and 24 that have some overlap in properties. In particular, we consider the GLEs with longitude and intensity similar to those of GLE72.

\subsection{Comparison with the solar cycle 24 GLE (GLE71)} \label{bozomath}
GLE71, almost a magnitude higher than GLE72, was associated with an M5.1 flare from W89 and a fast CME (${\sim}$1997 km/s). The CME was observed by STEREO-A/COR1, giving the CME height at SPR to be 2.32 Rs. The acceleration of the CME at the time of shock formation and SPR were 1.77 km/s${^2}$ and 1.51 km/s${^2}$ respectively \citep{Gopalswamy2013a}. We note that the CME speeds were comparable for GLE71 and GLE72 but the CME acceleration at SPR was slightly higher for GLE72. In addition, the GOES proton time profiles in Figure 1 are comparable to those in Figure 3(d) of Gopalswamy et al. (2013a), with similar intensity enhancements in the ${>}$700 MeV channel (${\sim}$84\% for GLE71 vs. ${\sim}$67\% for GLE72).

\subsection{Comparison with solar cycle 23 GLEs}
Here we compare the GLE72 CME heights at the m-Type II onset and SPR, and acceleration at SPR with those of the cycle 23 GLEs. Figure 4 shows the CME heights at the time of shock formation and SPR for GLEs of solar cycle 23 as a function of source longitude. The CME height at SPR (h$_s$ in Rs) shows a parabolic dependence on the source longitude (${\lambda}$ in degrees): h$_s$ = 2.55 + [(${\lambda}$-51)/35]${^2}$ \citep{Gopalswamy2012b}. Reames (2009a, b) obtained a similar figure using an independent technique. GLE72 lies ${\sim}$37\% below this curve.  We note that GLE65 (2003 Oct 28, longitude: E02) and GLE72 source longitude connectivity were similar. The GLE65 flare magnitude was X17, the most intense for all GLEs of cycles 23 and 24. CME speed and acceleration for GLE65 were 2754 km/s and 1.53 km/s${^2}$ respectively and the shock formed at a relatively low height of 1.27 Rs. The CME height at SPR is closer to the parabola for GLE65. The CME heights at SPR exhibit a standard deviation of ${\sim}$1.61 Rs from the parabola. The GLE72 CME is 1.71 Rs below the parabola but a few GLE CMEs deviate farther (e.g. GLE62 CME is ${\sim}$4.61 Rs higher).

In an extensive study of the CMEs associated with cycle 23 GLEs, Gopalswamy et al. (2012a) listed the mean and median CME accelerations to be 2.3 km/s${^2}$ and 2.02 km/s${^2}$ respectively. We note that the 2014 January 6 GLE CME has acceleration of 2.08 km/s${^2}$ at SPR and thus agrees with the acceleration of CMEs associated with cycle 23 GLEs.

\subsection{Comparison with a small GLE of solar cycle 23 (GLE68)}
Because the 2014 January 6 GLE was a small event with the maximum increase of ${\sim}$2.5\% , we compared it to the smallest GLE of cycle 23, viz., the GLE of 2005 January 17 (GLE68), for which the maximum increase of ${\sim}$3\% was observed. As shown superposed on Figure 4 (GLE68 marked in black), the CME heights at shock formation and SPR time were 1.44 Rs and 2.72 Rs respectively. The CME had a maximum speed of 2802 km/s and acceleration of 2.92 km/s${^2}$  (Gopalswamy et al. 2012a). These are higher than those for the GLE72 but the CME heights at shock formation and SPR are still consistent. It is interesting that although the GLE68 source location, W25, was within the well-connected solar region and the eruption was associated with a fast CME with a large acceleration, it produced a very small GLE. One of the possible reasons is the poor latitudinal connectivity for GLE68 \citep{Gopalswamy2013a}. Even though the flare location was N14W25, the flux-rope fit indicated a location of N18W25. The latitudinal connectivity became worse because of the unfavorable solar B0 angle (${\sim}$4.7${^{\circ}}$), making the effective location to be N23W25. The latitudinal distance to the ecliptic (23${^{\circ}}$) is 10${^{\circ}}$ greater than the nominal latitudinal distance (13${^{\circ}}$) for cycle-23 GLEs. The CME height at SPR falls on the parabola for GLE68.

\subsection{Comparison with a behind-the-limb GLE (GLE61)}
The only backside GLE of cycle 23 is the 2001 April 18 GLE (GLE61). The solar source was at W117. The associated CME had a speed of 2712 km/s and acceleration of 2.38 km/s${^2}$. In this event the time delay between the onset of the m-Type II and the SPR was about 15 minutes \citep{Gopalswamy2012a}. The time delay for GLE72 is about 10 minutes. As shown in Figure 4, the GLE61 CME was at heights of 1.47 Rs and 6.01 Rs at the m-Type II onset and the SPR time respectively. This is consistent with a particle event originating behind the limb, because the particle release occurs at a later time when the shock crosses the magnetically well-connected field lines. These two GLEs are on the right-side arm of the parabola because both occurred behind the west limb. GLE61 was ${\sim}$27${^{\circ}}$ behind the limb compared to only ${\sim}$12${^{\circ}}$ behind the limb for GLE72. Therefore, the CME height at SPR is smaller for GLE72, as expected. It is not clear if the smaller height at SPR for GLE72 has anything to do with the state of the heliosphere during a weak cycle.
The 2014 January 6 GLE, along with the 2001 April 18 GLE, forms a very interesting subset of GLEs that pose the greatest challenge to the flare acceleration mechanism for GLE particles. In the case of acceleration by flare, the particles would have been accelerated by a narrow source at the reconnection region, typically at the height of ${<}$0.05 Rs above the solar surface. Therefore, a flare acceleration region would not only be occulted but would also be poorly connected to Earth \citep{Gopalswamy2012a}, thus not resulting in a GLE. However, in both of these GLEs (61 and 72), the energetic particles were observed promptly at Earth. In the scenario of shock acceleration the CME-driven shock expands and the GLE particles can be observed when its eastern flank crosses the field lines that are magnetically connected to Earth. Therefore, flare accelerated particles would need to interact with the CME flux-rope to reach the well-connected field lines. In this case the first arriving particles will not remain scatter-free, contradicting the observed high anisotropy at the beginning of the GLEs. 



\section{Summary}
We have analyzed the GLE of 2014 January 6. Using STEREO-A/EUVI, STEREO-A/COR1 and SOHO/LASCO measurements to obtain accurate CME kinematics. GLE72 is unique in many aspects as it falls into different categories of a few special GLEs in cycles 23 and 24. In the unusual cycle 24, where there has been a dearth of GLEs, this is one of the only two GLEs. This is also one of the two smallest GLEs (GLE68 and GLE72) of cycles 23 and 24. In addition, GLE72 is one of the two behind-the-limb GLEs (GLE61 being the other one) of cycles 23 and 24, thus posing a challenge to flare acceleration theory for GLE particles. Our analysis for GLE72 is summarized as follows:

(i) The observation of ${>}$700 MeV protons by GOES confirms the 2014 January 6 to be a GLE, for which a small intensity enhancement was seen by the South Pole and a few other neutron monitors. This is consistent with findings by Gopalswamy et al. (2014) that all GLEs of cycle 23 and 24 had discernible enhancement in GOES ${>}$700 MeV proton channel.

(ii) The CME height at the time of shock formation (onset time of m-Type II radio emission) was measured directly by STEREO-A/EUVI to be 1.61 Rs. The CME height at SPR was measured directly by STEREO-A/COR1 to be 2.96 Rs. Our analysis is consistent with particle acceleration by a CME-driven shock. This is the first GLE for which STEREO provided direct CME height measurements both at the shock formation and SPR. This event compares well with the 2012 May 17 GLE in terms of GOES observations, CME heights at shock formation and SPR.

(iv) The CME height at the time of the shock formation is consistent with that of other GLE associated CMEs of cycle 23. The height of the CME at the SPR time varies with the longitude of the source of the GLEs such that the CME height at SPR increases as the source longitude goes beyond the best-connected regions (W20-80). The CME height at SPR for GLE72 is about 37\% below the parabola fitted to the cycle 23 GLEs. 

(v) The CME heights at the shock formation and SPR for the small GLE72 are comparable with those for the smallest GLE of cycle 23 (GLE68).

(vi) The CME height at SPR for GLE72 is consistent with that for GLE61; these GLEs originated 12${^{\circ}}$ and 27${^{\circ}}$ behind the limb respectively.

\acknowledgments
We thank Dr. P. Evenson of Bartol Neutron Monitor Network for the neutron monitor asymptotic directions and useful discussions. We thank Dr. T. Kuwabara of Bartol for his help with conversion of the IMF directions for Figure 3. We acknowledge the NMDB database (www.nmdb.eu), founded under the European Union's FP7 programme (contract no. 213007) for providing data. Neutron monitors of the Bartol Research Institute are supported by the National Science Foundation. STEREO is a mission in NASA's Solar Terrestrial Probes program. SOHO is a project of international collaboration between ESA and NASA. This work was supported by NASA's LWS TR\&T program.

\clearpage

\begin{figure}
\epsscale{.80}

\includegraphics[height=210pt]{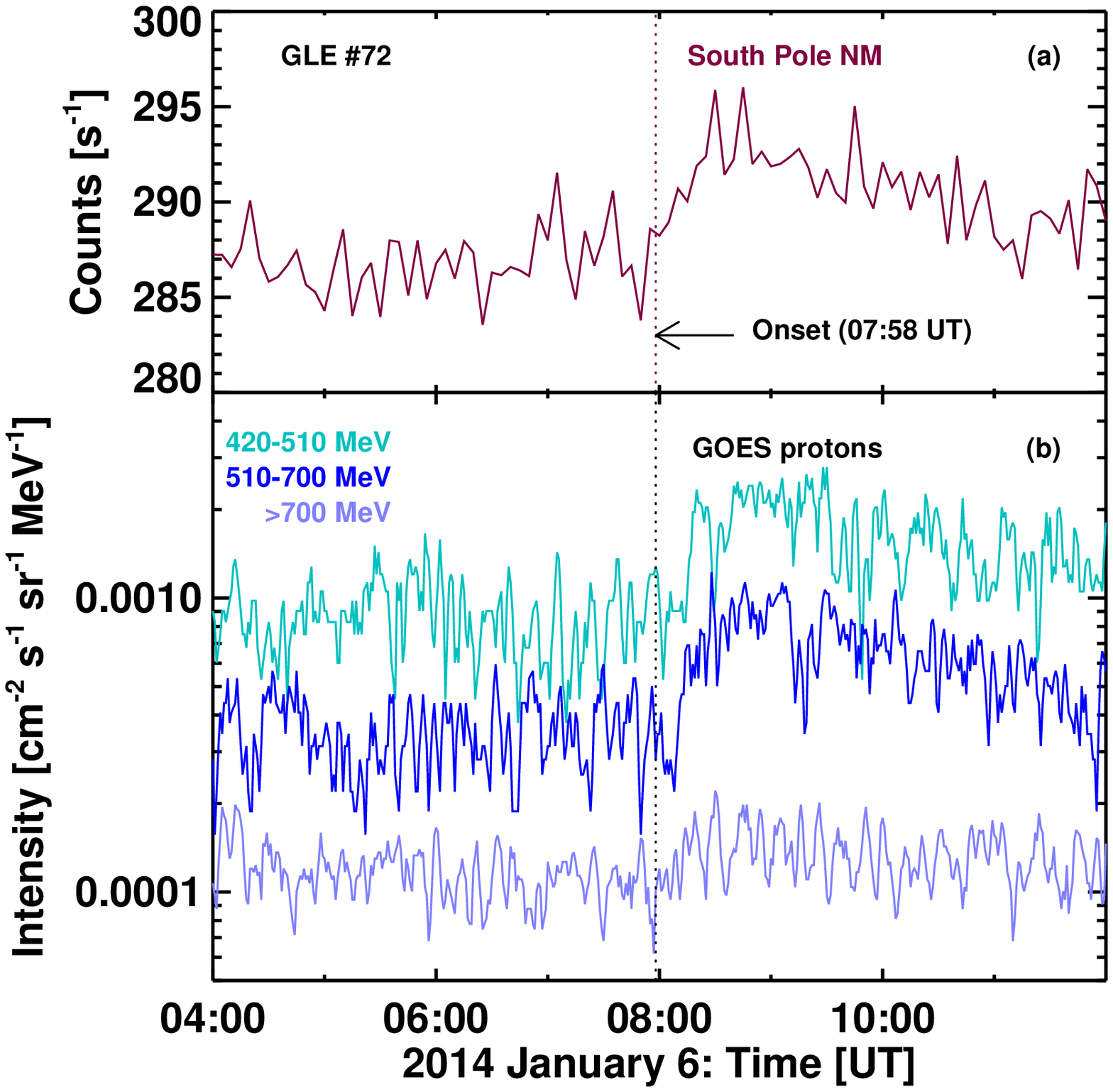}
\includegraphics[height=200pt]{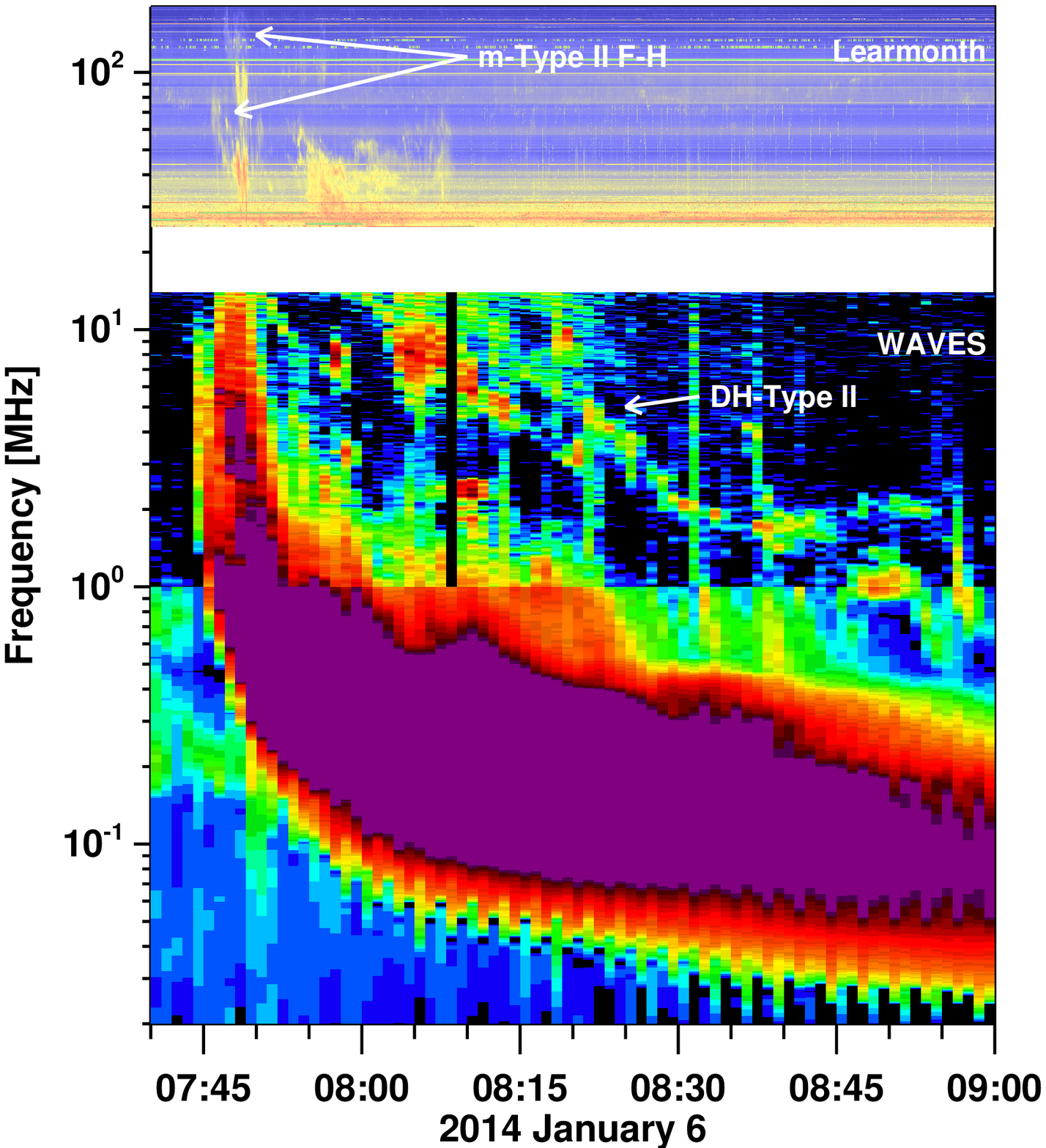}

\caption{Left: (a) Count rates from the South Pole Neutron Monitor (http://neutronm.bartol.udel.edu). A rise of ${\sim}$2.5\% was observed. (b) Proton intensity time profile from GOES/EPEAD. GOES observed ${\sim}$67\% increase in ${>}$700 MeV channel. The onset at SPNM (dotted vertical line) occurs before GOES. The normalized Solar Particle Release time was 07:55 UT. Right: (upper) m-Type II radio burst observed by the Learmonth radio observatory (www.ips.gov.au/learmonth) starting at 07:45 UT (the arrows point to the fundamental-harmonic (F-H) bands) and (lower) the decameter-hectometric (DH; pointed by an arrow) to kilometric type II observed by Wind/WAVES. 
\label{fig1}}
\end{figure}

\clearpage

\begin{figure}
\includegraphics[height=210pt]{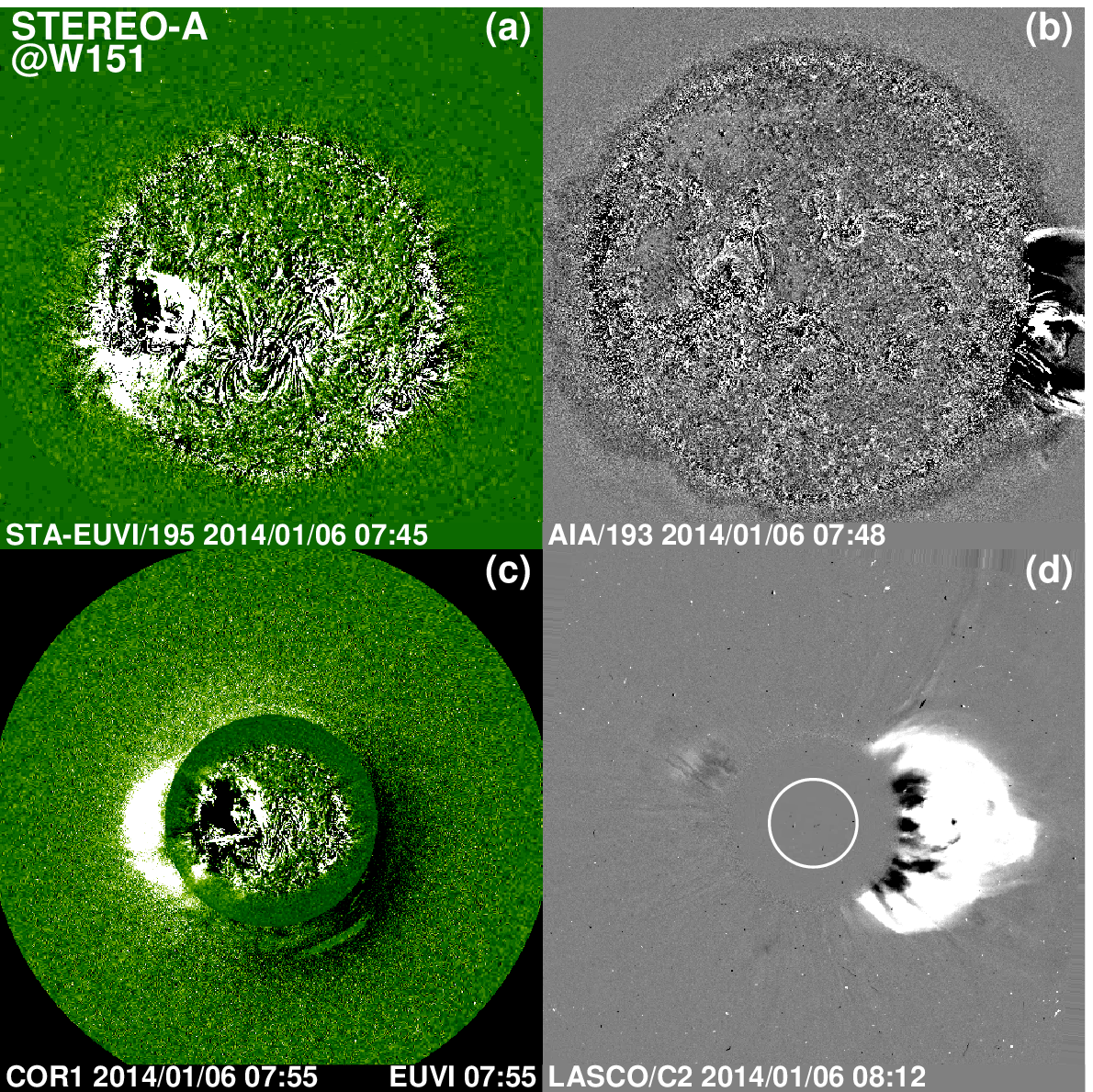}
\includegraphics[height=220pt]{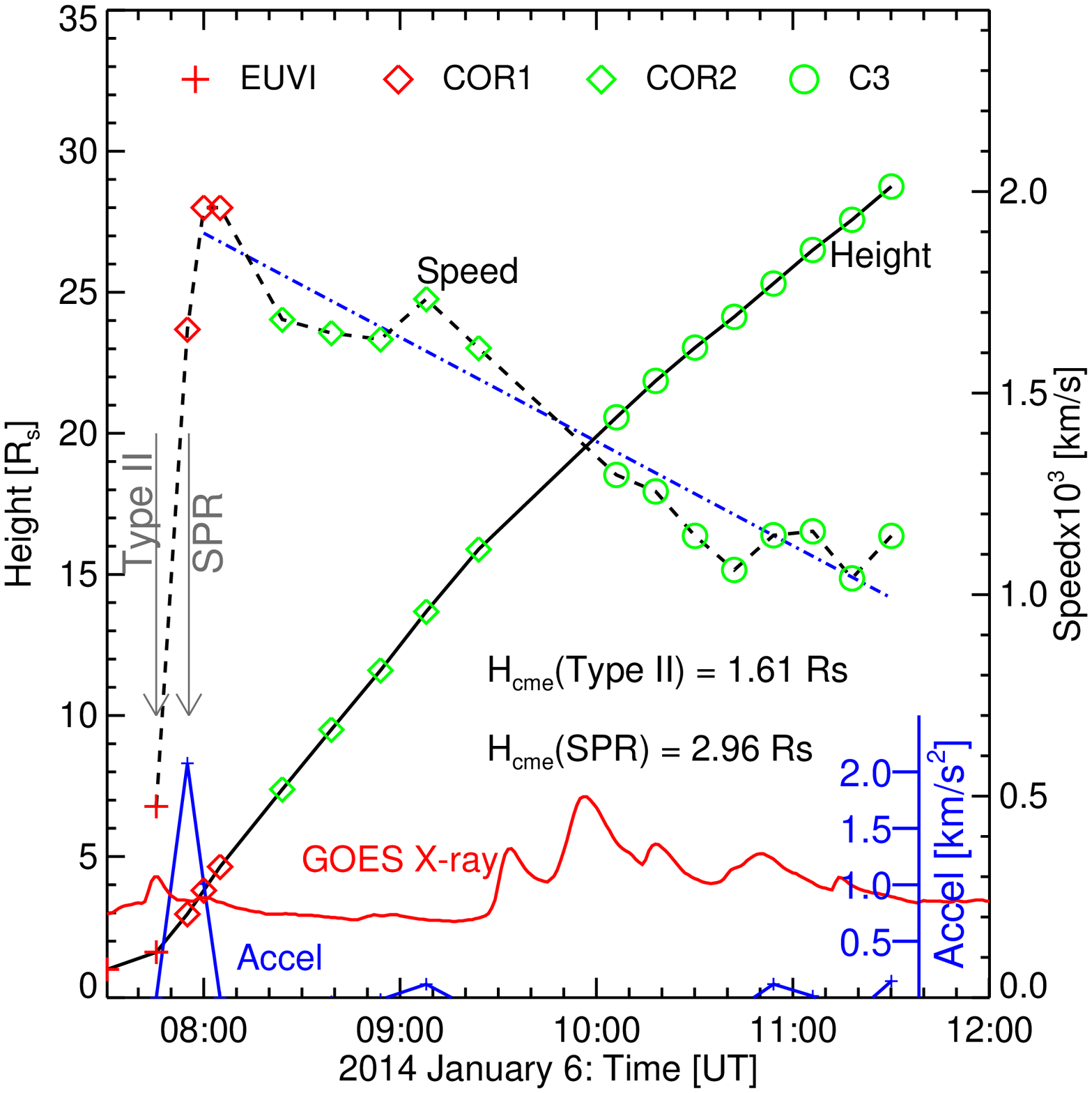}
\caption{Left: (a) The 2014 January 6 CME observed by STEREO-A/EUVI at the onset of m-Type II (07:45 UT) burst. (b) The post-flare loops seen above the west limb in the SDO/AIA 193 \AA\ image (07:48 UT). (c) STEREO-A/COR1 and EUVI image of the CME at SPR (07:55 UT). (d) The CME as seen by LASCO/C2 (08:12 UT). Right: The CME kinematics obtained by fitting a flux-rope to STEREO-A/EUVI, COR1, COR2, and SOHO/LASCO observations. The height and acceleration of the CME were ${\sim}$2.96 Rs and 2.08 km/s${^2}$ respectively at SPR. The CME speed at SPR was 1658 km/s. The GOES 1.0-8.0 \AA\ soft X-ray flux is superposed.
\label{fig2}}
\end{figure}

\begin{figure}
\plotone{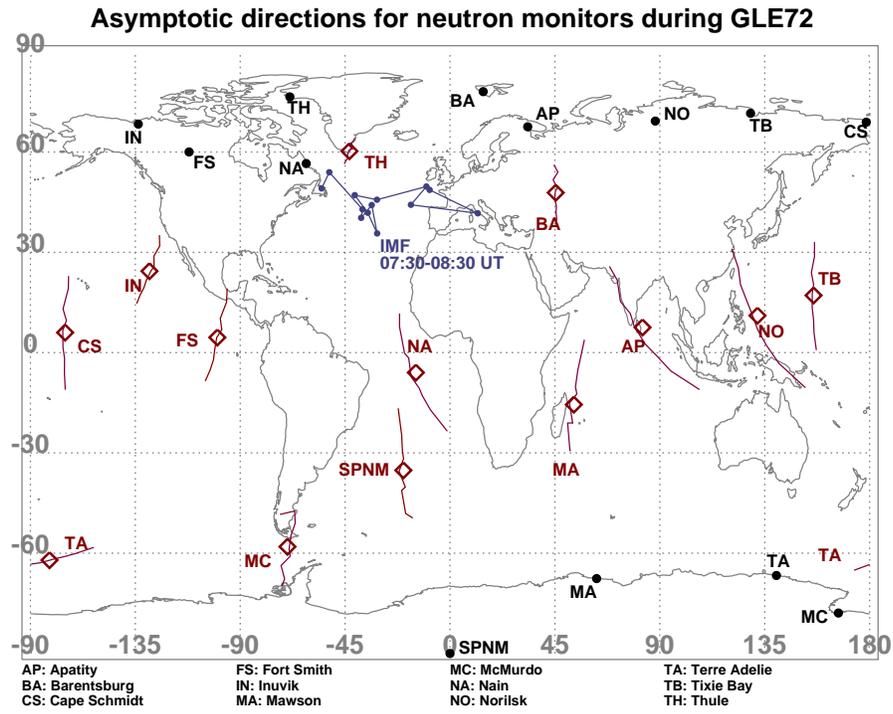}
\caption{Neutron monitor viewing directions (red) at 07:58 UT on 2014 January 6. The particle arrival directions based on the IMF direction are also shown (blue).
\label{fig3}}
\end{figure}

\begin{figure}
\plotone{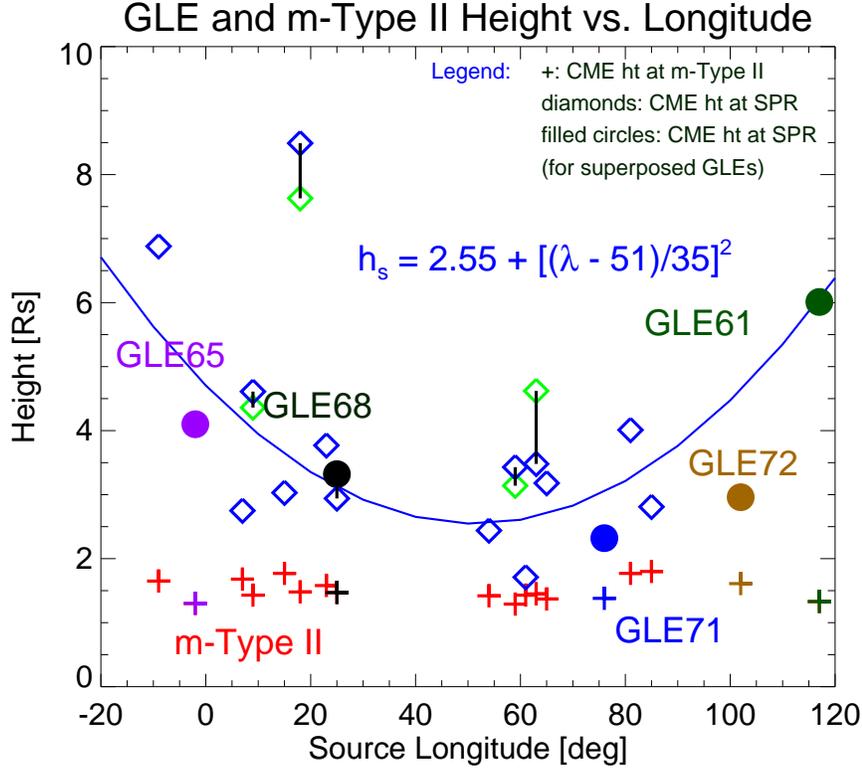}
\caption{The height of the CME at shock formation and SPR plotted against the source longitude of cycle 23 GLEs. The CME heights at SPR can be fitted with a parabola as shown. The 2014 January 6 event (GLE72) as well as GLEs61, 65, 68, and 71 are shown superposed. The green and blue diamonds indicate heights determined from linear and quadratic extrapolations. 
\label{fig4}}
\end{figure}

\clearpage

\begin{deluxetable}{lr}
\tabletypesize{\scriptsize}
\tablecaption{Timeline of events for GLE72. All times are Earth times and propagation times for electromagnetic emissions have been considered in this table as explained in the text.}
\tablewidth{0pt}
\tablehead{
\colhead{Event} & \colhead{Time [UT]}
}
\startdata
CME onset & 07:30\\
CME first observation by STEREO-A/EUVI & 07:30\\
Metric type II onset & 07:45 \\
Normalized SPR time & 07:55 \\
CME first observation by STEREO-A/COR1 & 07:55 \\
GLE particles arrival at SPNM & 07:58 \\
CME first observation by SOHO/LASCO & 08:00 \\
\enddata
\end{deluxetable}


\end{document}